\documentclass[twocolumn,a4paper,english,aps,prl,superscriptaddress]{revtex4}
\usepackage[latin9]{inputenc}
\usepackage{amsmath}
\usepackage{amssymb}
\usepackage{esint}
\usepackage{epsfig}
\usepackage{epstopdf}
\usepackage{bm}
\usepackage[usenames]{color}
\usepackage{babel}

\makeatletter
\@ifundefined{textcolor}{}
{
 \definecolor{BLACK}{gray}{0}
 \definecolor{WHITE}{gray}{1}
 \definecolor{RED}{rgb}{1,0,0}
 \definecolor{DARKYELLOW}{rgb}{1,0,0}
 \definecolor{GREEN}{rgb}{0,1,0}
 \definecolor{BLUE}{rgb}{0,0,1}
 \definecolor{olive}{rgb}{0.2,0.5,0}
 \definecolor{CYAN}{cmyk}{1,0,0,0}
 \definecolor{MAGENTA}{cmyk}{0,1,0,0}
 \definecolor{YELLOW}{cmyk}{0,0,1,0}
}
\makeatother

\begin{document}

\title{Anomalous Hall effect in a 2D magnetic semiconductor hetereostructure: Evidence of Berry phase effects in the hopping-transport regime}

\author{L.~N.~Oveshnikov}
\affiliation{National Research Center Kurchatov Institute, Moscow 123182, Russia}
\author{V.~A.~Kulbachinskii}
\affiliation{National Research Center Kurchatov Institute, Moscow 123182, Russia}
\affiliation{Low Temperature Physics Department, M.V.~Lomonosov Moscow State University, Moscow 119991, Russia}
\author{A.~B.~Davydov}
\affiliation{P.N.~Lebedev Physical Institute, Russian Acad. Sci., Moscow 119991, Russia}
\author{B.~A.~Aronzon}
\affiliation{National Research Center Kurchatov Institute, Moscow 123182, Russia}
\affiliation{P.N.~Lebedev Physical Institute, Russian Acad. Sci., Moscow 119991, Russia}
\author{I.~V.~Rozhansky}
\affiliation{Ioffe Institute, Russian Acad. Sci., St.~Petersburg 194021, Russia}
\affiliation{Peter the Great St.~Petersburg Polytechnic University, St.~Petersburg 195251, Russia}
\author{N.~S.~Avkeriev}
\affiliation{Ioffe Institute, Russian Acad. Sci., St.~Petersburg 194021, Russia}
\author{K.~I.~Kugel}
\affiliation{Institute for Theoretical and Applied Electrodynamics, Russian Acad. Sci., Moscow 125412, Russia}
\author{V.~Tripathi}
\affiliation{Materials Science Division, Argonne National Laboratory, Lemont, IL 60439, USA}
\affiliation{Department of Theoretical Physics, Tata Institute of Fundamental Research, Mumbai 400005, India}

\begin{abstract}
The temperature and magnetic field dependences of the anomalous Hall effect (AHE) are studied in Mn delta-doped semiconductor heterostructures 
with a two-dimensional hole gas in a quantum well interacting with the Mn layer. The analysis of experimental data reveals four distinct temperature 
ranges differing in the behavior of AHE. The Mn layer induces an inhomogeneity of the hole gas leading to an interplay between the hopping and drift 
conductivities. It is shown that at sufficiently low temperatures,
hopping conductivity favors the mechanisms of AHE related to the geometric (Berry--Pancharatnam) phase.
\end{abstract}

\maketitle


The anomalous Hall effect (AHE) is widely observed in ferromagnets in the presence of spin-orbit interaction~\cite{nagaosa2010}. It accurately indicates 
the onset of spin polarization of the carriers -- a key requirement for spintronic devices. Existing theories of AHE in semiconductors based on the 
consideration of different spin-dependent scattering processes \cite{smit1955,berger1964,berger1970}, or geometric (Berry--Pancharatnam) phase 
effects~\cite{onoda2002,jungwirth2002} have been quite successful in describing AHE in the metallic regime (such as in Mn-doped GaAs). Much less 
understood is AHE in the hopping transport regime, and while there are some proposals for this regime~\cite{burkov2003,lyanda2001}, they are inadequately 
examined in experiments. Crucially, all these studies deal with AHE in bulk systems. The paucity of AHE literature on two-dimensional (2D) semiconductors 
seems quite surprising given the considerable current interest in spintronics and the fact that modern semiconductor technology is essentially planar. 
A parallel motivation to study AHE in 2D semiconductors stems from the recent discovery~\cite{chang2013} of quantized AHE on 2D surfaces of 3D topological 
insulators in the presence of a small concentration of magnetic impurities that undergo ferromagnetic ordering. The quantized AHE is the result of 
two-dimensionality and strong spin-orbit coupling in the topological insulator and is directly related to the quantization of the geometric flux accumulated by 
Bloch states around the Brillouin zone, analogous to the quantized ordinary Hall effect~\cite{thouless1982}. Likewise, a novel thermal Hall effect recently
reported~\cite{ong} in insulating bosonic 2D spin-liquids is in fact an AHE associated with geometric phase effects.

In the hopping regime, while Bloch states no longer exist, the ordinary Hall effect now arises from the Aharonov--Bohm phase accumulated by 
elementary triangles~\cite{holstein1961} on the percolating backbone, and is thus a quantum phase effect. It is then an interesting open question 
whether a geometric phase effect is similarly responsible for AHE in 2D magnetic semiconductors in the hopping regime. Previous experimental 
studies~\cite{allen2004} of 3D magnetic semiconductors have argued against this possibility. In this Letter, we address this issue in our experimental 
study of AHE in 2D magnetic semiconductor quantum wells with (hole) charge transport in the hopping regime. Our main finding concerns an evidence of 
the geometrical phase related mechanism for AHE at sufficiently low temperatures. We also find that at higher temperatures, 
AHE is governed by an alternative mechanism based on a spin-dependent hopping of 2D holes in QW interacting with remote magnetic layer, 
reminiscent of mechanism suggested in Ref.~\onlinecite{burkov2003} and observed~\cite{allen2004} for bulk magnetic semiconductors.

GaAs/$\delta-$Mn/GaAs/In$_{x}$Ga$_{1-x}$As quantum well (QW) structures
were grown by MOCVD methods on semi-insulating GaAs substrates, using laser
deposition of Mn. Two of the samples, M1 and M2, were cut from the same wafer.
The main part of each sample is the In$_{x}$Ga$_{1-x}$As QW sandwiched between two GaAs layers. One of them acts as a 3 nm spacer separating the QW
from the Mn $\delta$-layer with the effective Mn concentration of 0.3 monolayer. Our previous X-ray studies~\cite{aronzon2008} have shown that the 
composition of this layer is indeed Ga$_{1-x}$Mn$_{x}$As with Mn distributed
over a range of 2 nm, and around 5\% Mn-doping at the center of the layer giving a QW--Mn separation of 2--2.5 nm. Bulk GaAs films with such a
concentration of Mn are usually ferromagnetic with the Curie temperature slightly above 100 K. Mn in GaAs and in our structures plays a dual role as an
acceptor and a magnetic impurity. Indium content $x$ in the strained channel layer In$_{x}$Ga$_{1-x}$As is 0.216 for both samples (see Ref.~\cite{oveshnikov2014}
for a detailed description of the sample structure). The magnetic field and temperature dependences of longitudinal and Hall resistivities were analyzed in the
temperature range 2--80K, and for magnetic fields up to 8 K. We find that the hole concentration $p$ obtained from the Hall measurements is almost temperature
independent below 35--40 K (with $p\approx0.4\times10^{12}\mathrm{cm}^{-2}$).
\begin{figure}[tbp]
\centerline{\psfig{figure=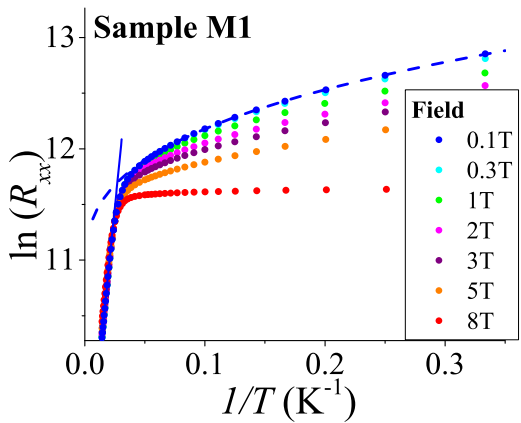,width=0.8\columnwidth}}
\caption{(Color online) Temperature dependence of the longitudinal
conductivity for sample M1 for different values of perpendicular magnetic
field $B.$ The Arrhenius law at higher temperatures (straight line fit)
gives way to a 2D Mott variable range hopping law at lower temperatures
(dashed line fit).}
\label{fig:VRH}
\end{figure}
The spatial distribution of Mn atoms within the impurity layer is
inhomogeneous \cite{aronzon2010}, which creates fluctuation potentials and phase separation in the impurity (Mn) layer as well as
the transport channel (QW)
\cite{aronzon2008}.
Parameters of the fluctuation potential and its
effect on the heterostructure conductivity and magnetic properties were
discussed in Refs. \onlinecite{tripathi2011,tripathi2012}.

Fig.~\ref{fig:VRH} shows the temperature dependence of the longitudinal resistivity for sample M1 for different
values of perpendicular magnetic field $B.$ A transition from activated to
variable-range hopping (VRH) conduction is visible as the temperature is lowered. In the hopping regime, a better
fit is achieved by the 2D Mott VRH law $\ln R_{xx}\sim(1/T)^{1/3}$ compared to the Efros--Shklovskii VRH $\ln R_{xx}\sim(1/T)^{1/2}$
signifying that long-range Coulomb interactions are not important in this temperature range. From the intersection of
Arrhenius and VRH fitting curves, we estimate the
temperature $T_{\text{hop}}$ at which the transition from drift to hopping
transport occurs. For sample M1, we find $T_{\text{hop}}=37$ K ($B=0.1$ T) and 38.5 K ($B=8$ T).
\begin{figure}[tbp]
\centerline{\psfig{figure=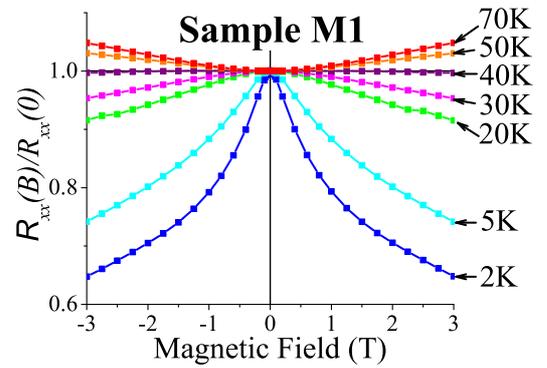,width=0.8\columnwidth}}
\caption{(Color online) Magnetic field dependence of $R_{xx}(B)/R_{xx}(0)$
at different temperatures. As the temperature is decreased, the curvature
changes sign around $T=40$K. It is argued in the text that the negative
sign at low temperatures is not a quantum interference effect but rather a
sign of increasing ferromagnetic correlations.}
\label{fig:RxxB}
\end{figure}

Fig.~\ref{fig:RxxB} shows the results of magnetoresistance measurements. Upon lowering the temperature, a transition from positive to negative magnetoresistance
occurs around $T=40$ K, which incidentally is very close to $T_{\text{hop}}$.
We attribute the positive magnetoresistance to the localizing (orbital) effect of the magnetic field in the drift regime.
In the hopping regime, an opposing contribution to the magnetoresistance comes from the dependence of the hopping rate on the
alignment of the magnetic moments. We thus believe that the negative magnetoresistance is related to spin-dependent tunneling and an indication
of the trend to onset of ferromagnetic order
rather than a quantum interference effect (see, e.g. Refs.~\cite{tripathi2011,tripathi2006}).

The Mn impurity layer determines the magnetic properties. Exchange interaction mediated by itinerant carriers
in the (Ga,Mn)As layer~\cite{menshov2009} and free carriers in the QW \cite{meilikhov2008,OurArxiv2015}
lead to ferromagnetic (FM) ordering of the Mn magnetic moments. The Mn-layer is phase separated into ferromagnetic
regions (with a high Mn content) inside a non-ferromagnetic
matrix, and undergoes a percolation transition to a long-range ferromagnetic state at low temperatures. In these 2D
disordered magnets (unlike 3D counterparts), the Curie temperature, at which local ferromagnetic order appears
(as detected through AHE measurements) and the temperature $T_{c}$,
at which ferromagnetic clusters begin percolating (visible as a peak in the temperature dependence
of $\ dR_{xx}/dT$) can be rather far apart~\cite{tripathi2011}.
The structure demonstrates ferromagnetic ordering \cite{aronzon2010,tripathi2011} with $T_{c}=24K$
estimated from the peak in $dR_{xx}(T)/dT$ vs $T$
~\cite{tripathi2011,wang2014}.

\begin{figure}[tbp]
\centerline{\psfig{figure=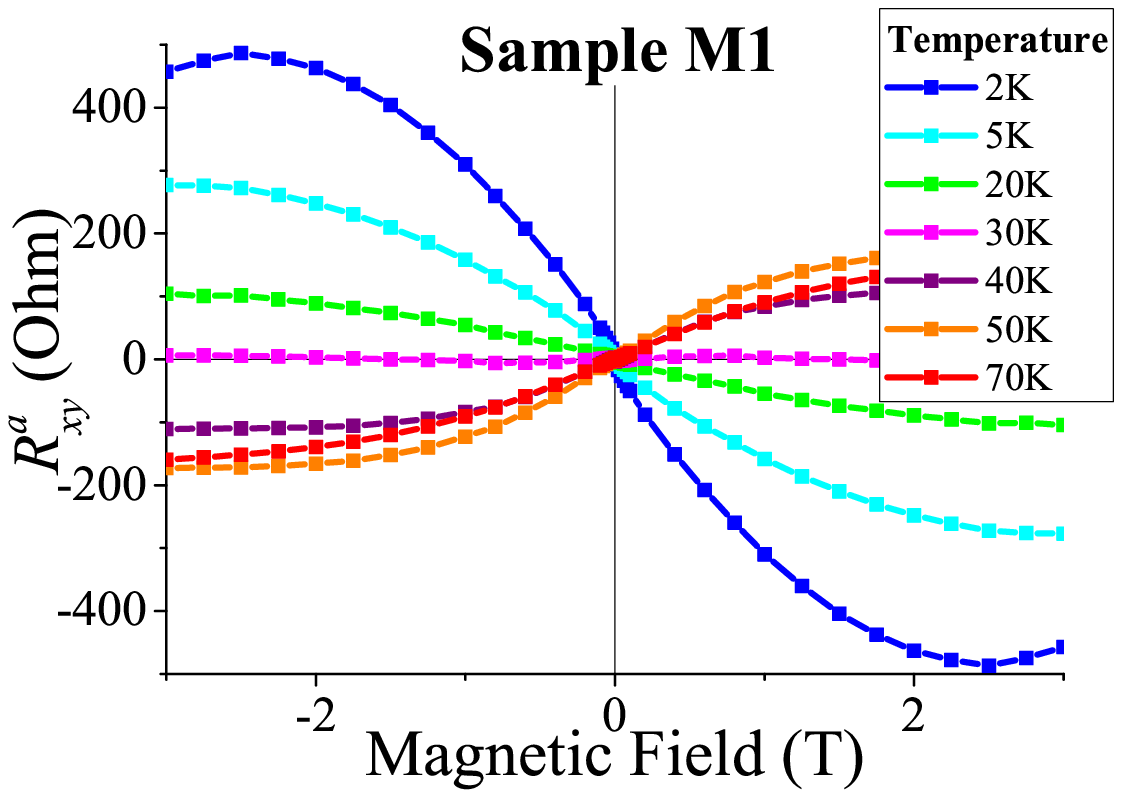,width=0.8\columnwidth}} \centerline{%
\psfig{figure=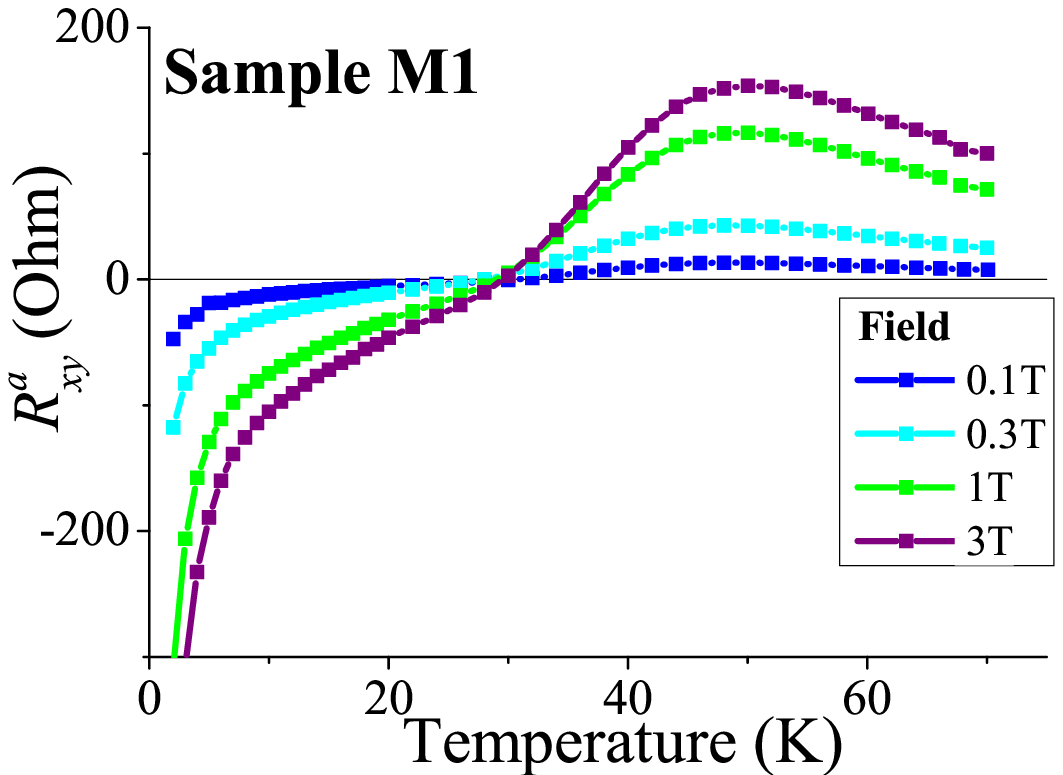,width=0.8\columnwidth}}
\caption{Magnetic field and temperature dependences of the anomalous Hall
resistance $R_{xy}^{a}$ of sample M1. Saturation is observed above fields of
around 2T. The anomalous Hall resistance increases with temperature over a
considerable range - a surprising feature also observed in two-dimensional
thin films of GaMnAs \protect\cite{chiba2010}. $R_{xy}^{a}$ also changes
sign around $T_{c}.$}
\label{fig:Rxy-BT}
\end{figure}

The AHE contribution is obtained from the full Hall signal by separating the
linear-in-$B$ part, which is possible if the sample magnetization
saturates. We show the magnetic field and temperature dependences of the anomalous Hall
resistivity $R^{a}_{xy}$ we have obtained in Fig.~\ref{fig:Rxy-BT}.
The field dependences typically tend to saturate above $2\mathrm{T}$ (see also Refs. \cite{aronzon2010,aronzon2008,tripathi2011}), although
the lowest temperature curve shows a small downturn in
$|R^{a}_{xy}|$ instead of saturation. This can arise, for example, if there are two contributions to the conductivity in the QW
(the hopping between metallic droplets and drift conductivity inside them).
One striking feature of the observed AHE is the increase of $R_{xy}^{a}$
with temperature, very similar to the observation in Ref.~\cite{chiba2010} in thin GaMnAs layers. Reduced dimensionality, a common feature of the
two systems, may be the cause of such anomalous behavior. Note also that in our structures, $R_{xy}^{a}$ changes sign upon cooling around $T_c$,
the magnetic percolation temperature.

\begin{figure}[tbp]
\centerline{\psfig{figure=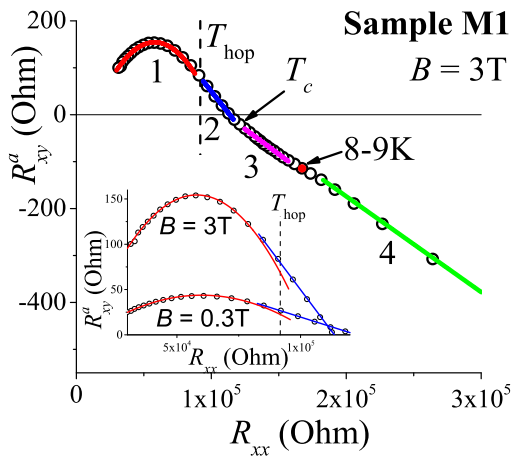,width=0.8\columnwidth}}
\caption{(Color online) Parametric dependence of anomalous Hall resistivity $R_{xy}^{a}$ on longitudinal resistivity $R_{xx}$ for $B=3$ T fitted by a
linear function at low temperatures and by a quadratic function at higher
temperatures. Inset shows linear-to-quadratic behavior change around $T_{\text{hop}}$.}
\label{fig:balents-mech}
\end{figure}

Different physical mechanisms of AHE may become relevant as the parameters are varied. In bulk metals, the scaling of $R_{xy}^{a}$ with
longitudinal resistivity is commonly used to determine the physical mechanism. A linear dependence $R_{xy}^{a}\propto R_{xx}$ is a signature of the
scattering-dependent skew-symmetric mechanism, while a quadratic dependence
$R_{xy}^{a}\propto R_{xx}^{2}$ indicates the side-jump or intrinsic mechanisms, the latter being scattering-independent. The hopping transport regime
needs a substantially different theoretical approach.
In Ref.~\onlinecite{burkov2003}, a mechanism of AHE in a 3D disordered ferromagnet in the hopping regime was suggested based on spin-dependent hopping between the sites
as a consequence of spin-orbit interaction and spin polarization of the carriers in effective exchange mean field.
The theory predicts a linear parametric dependence of $R_{xy}^{a}$ on $R_{xx},$
$R_{xy}^{a}\sim \rho _{0}^{\prime }(\epsilon _{F})R_{xx},$
where $\rho _{0}$ is the density of states at the Fermi level. The sign of AHE relative to the ordinary Hall effect may vary depending on the
position of the Fermi level within the impurity band. An experimental observation of both signs of AHE in (Ga,Mn)As hopping transport was reported and
interpreted using this theory \cite{allen2004}. A different model for the AHE in the hopping regime was developed for manganites \cite{lyanda2001},
where AHE arises from a Berry--Pancharatnam geometric phase due to misalignment of local magnetic moment
axes and thus vanishes as the magnetization saturates. It was claimed in Ref.\cite{allen2004} that the
geometric phase contribution is not relevant in magnetic semiconductors on the basis of their observation that AHE
does not vanish at saturation magnetization. This argument however is correct only if a single mechanism dictates AHE.
Below we make a careful analysis of the field dependence of AHE and argue that in our experiments
we observe both hopping AHE mechanisms mentioned above
along with the AHE reminiscent of side jump or intrinsic mechanisms in the band conductivity regime.
Fig.~\ref{fig:balents-mech} shows anomalous Hall data for sample M1 at $B=3$ T.
We observed four temperature ranges with different parametric
dependences $R_{xy}(R_{xx})$ corresponding to (with decreasing temperature)
the crossover to hopping transport at $T_{\text{hop}}$, ferromagnetic ordering at $T_c$, and to the puzzling appearance of a new AHE
mechanism at low temperature which we describe below.
At higher temperatures, where conductivity is determined by drift transport of holes at the percolation level, we find the sign of AHE is
positive relative to the ordinary Hall effect and is described by a quadratic fit, $R_{xy}^{a}=R_{0}+A(R_{xx}-R_{c})^{2}.$
This is reminiscent of scattering-independent mechanisms seen~\cite{nagaosa2010} in metallic (Ga,Mn)As. A linear relation $R_{xy}^{a}\propto R_{xx}$
provides a good fit to the data below the hopping transition temperature $T_{\text{hop}}$. Here, we observe a decrease of $R_{xy}^{a}$ with lowering $T$
(and rising $R_{xx}$).
At $T_c\approx 24$ K, corresponding to the magnetic percolation transition, the slope $dR_{xy}^{a}/dR_{xx}$ changes together with the sign of $R_{xy}^{a}.$
The low-temperature regime in Fig.~\ref{fig:balents-mech} has another peculiarity - an inflection point corresponding to a temperature of 8--9K.
At this point, the second derivative becomes zero, while the first derivative has a peak. We argue that in this lowest temperature range
the AHE is governed by the Berry--Pancharatnam geometric phase mechanism for the first time observed in two-dimensional system.

\begin{figure}[tbp]
\centerline{\psfig{figure=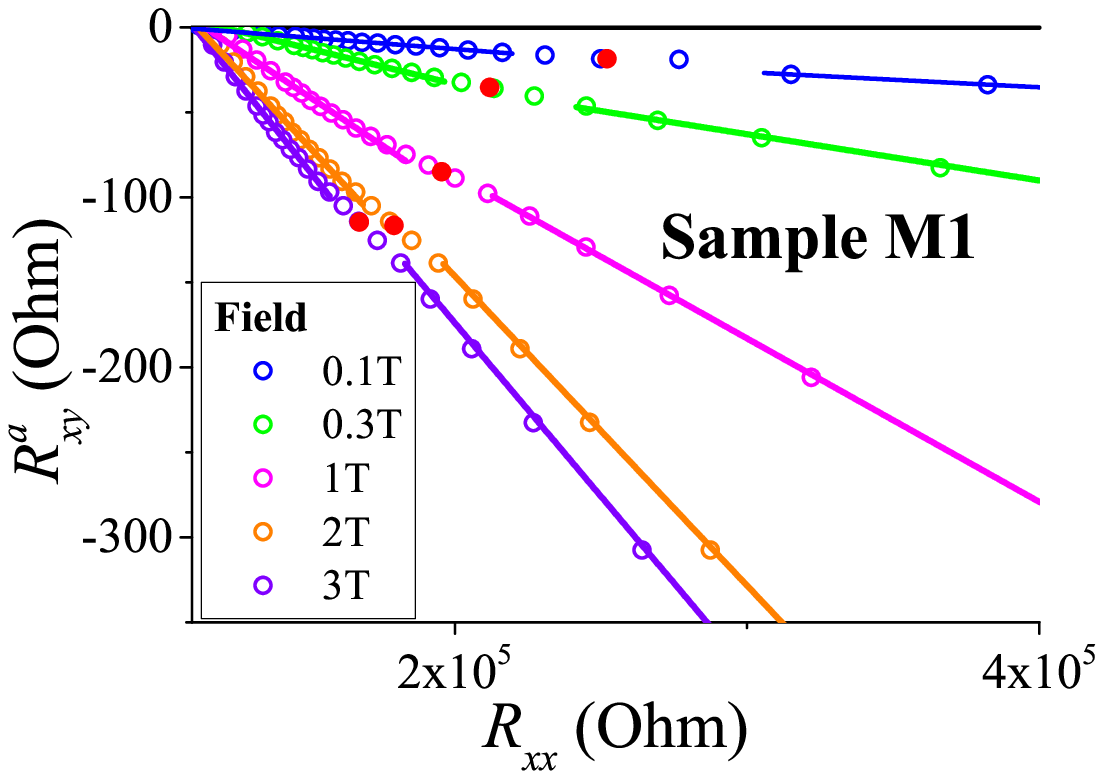,width=0.8\columnwidth} }\centerline{\psfig{figure=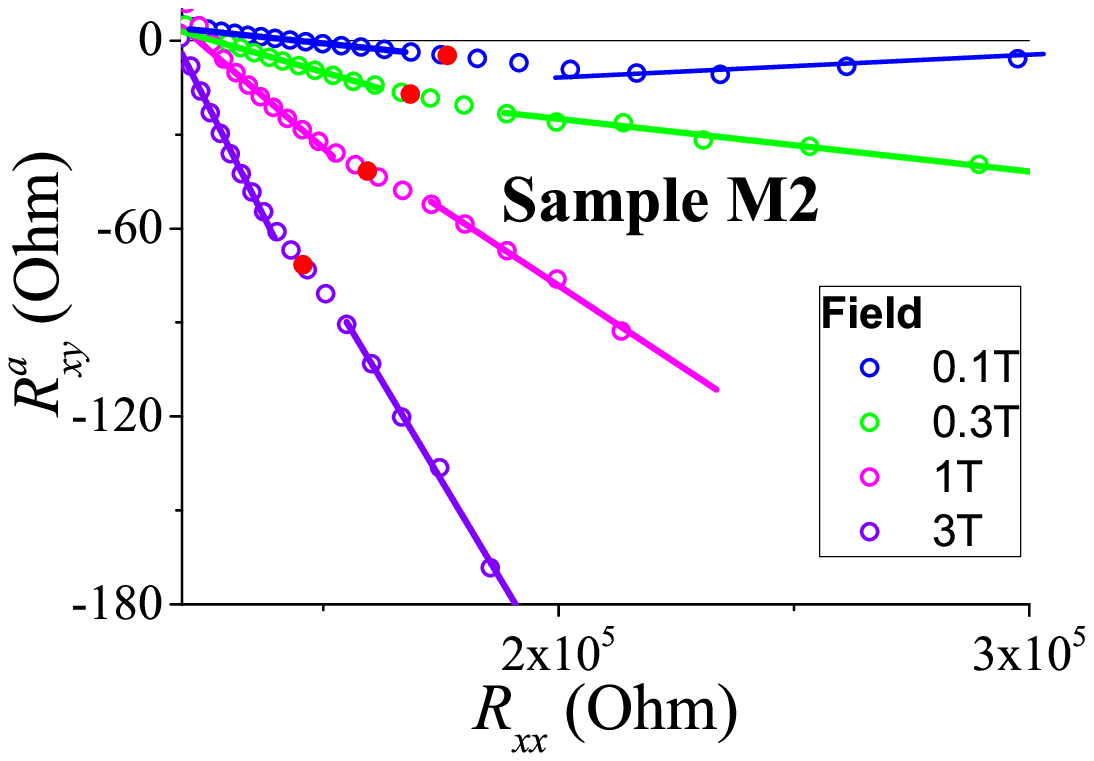,width=0.8\columnwidth}}
\caption{(Color online) Parametric dependences ($R_{xy}^{a}$ vs. $R_{xx}$)
at low temperatures for different values of the magnetic field. The behavior
of $R_{xy}^{a}$ below the inflection points (shown as red dots) on the
curves shows strong sensitivity to magnetic fields as small as 0.1 T,
and this effect goes away at higher fields $\sim3$T, where the impurity
magnetization is saturated. This contribution is opposite in sign compared
to the high field trend and even manifests as an upturn in sample M2. An
interpretation in terms of a geometric phase effect is given in the text.}
\label{fig:berry}
\end{figure}

First, let us summarize the emerging physical picture for AHE in our system in the temperature regions I--III. 
We have two contributions to conductivity: (a) hopping between metallic droplets and (b) drift conductivity inside them and at percolation level. 
At high temperatures (Region I), drift conductivity prevails, AHE is positive and a quadratic parametric dependence of  $R_{xy}^a(R_{xx})$ fits 
experimental results (see the lower inset of Fig.~\ref{fig:balents-mech}). In this region the AHE is governed by the dissipationless mechanism, 
reminiscent of side-jump or intrnsic, while the current is carried by the delocalized band states.
With rising magnetic field, the magnetization increases resulting in enhancement of AHE.
Now let us discuss AHE in the hopping regime. It is instructive to recall the mechanism of the ordinary Hall effect in the hopping regime.
As was first shown by Holstein~\cite{holstein1961}, the Hall conductivity $\sigma_{xy}$ in disordered insulators is determined by the Aharanov--Bohm
flux $\Phi =\int_{\vartriangle }\mathbf{A}\cdot d\mathbf{l}$
threading elementary triangles on the conducting backbone:
$\sigma _{xy}\sim \text{Im}(t_{12}t_{23}t_{31})\sin (2\pi \Phi /\Phi _{0}),$
where $t_{ij}$ denote the hopping matrix elements between sites $i$ and $j$, and
in our case metallic droplets inside the QW.
The presence of spin-orbit interaction makes hopping spin-dependent, and together with the spin polarization
of the holes in a exchange mean field of Mn ions, one gets AHE exactly as described in \cite{burkov2003}.
In Region II between $T_{\text{hop}}$ and $T_c,$ hopping transport starts to dominate over drift band conductivity.
We assume in our case the hopping contribution to AHE is of the opposite sign to the band
conductivity contribution.
In the region II, however, while the hopping conductivity dominates over drift the drift contribution to AHE
is still stronger and $R^{a}_{xy}$ maintains a positive sign.
Physically, this is possible as hopping contributes less to the Hall effect
than it does to the conductivity. With further decreasing temperature, the drift contribution to AHE is eventually superseded by the hopping one.
  The sudden enhancement of spin-dependent tunneling at the magnetic percolation temperature $T_c$ strongly enhances the hopping contribution and causes a
sign change in $R^{a}_{xy}.$ Further, since the establishment of long-range ferromagnetic order weakens the temperature dependence of
magnetization (and $R_{xy}^a$), the slope of $R_{xy}^a(R_{xx})$ diminishes upon crossing $T_c.$
Fig.~\ref{fig:berry} shows the evolution of parametric dependences for both samples in different magnetic fields. Since the drift 
and spin-dependent hopping contributions increase with magnetic field (through its effect on the magnetization) in the entire temperature 
range, the gradient of $R^{a}_{xy}(R_{xx})$ is expected to increase with $B$. However, in Region III below $T_{c},$ the dependence of the gradient 
of $R_{xy}^a(R_{xx})$ on magnetic field grows weaker.

Finally, we discuss AHE at very low temperatures, below the temperature of the inflection point (Region IV). The different roles played by the spin-orbit 
interaction (SOI) in the hopping conductivity need to be appreciated here. First, it generates AHE by imparting a spin-dependence to the hopping of the 
spin-split carriers between localized sites in the QW. This contribution exists over the entire temperature range
provided there is a spin polarization of the carriers due to the exchange mean field.
This contribution is normally regarded as the main cause for AHE in the hopping regime discussed in the existing literature for bulk
GaMnAs \cite{Jungwirth2006,burkov2003,allen2004}. We also belive this mechanism to be of key importance for the temperature regions II and III.
The second role of SOI is that it generates a Dzyaloshinskii--Moriya (DM) interaction of magnetic moments on a triad. In this case as a carrier hops around 
this triangle, it picks an additional geometric (Berry--Pancharatnam) phase $\Omega /2$, which is the quantum phase associated with the 
wavefunction overlaps $\langle \mathbf{n}_{1}|%
\mathbf{n}_{2}\rangle \langle \mathbf{n}_{2}|\mathbf{n}_{3}\rangle \langle
\mathbf{n}_{3}|\mathbf{n}_{1}\rangle ,$ $\tan (\Omega /2)=(\mathbf{n}%
_{1}\cdot \mathbf{n}_{2}\times \mathbf{n}_{3})/(1+\mathbf{n}_{1}\cdot
\mathbf{n}_{2}+\mathbf{n}_{2}\cdot \mathbf{n}_{3}+\mathbf{n}_{3}\cdot
\mathbf{n}_{1}).$ This geometric phase, essentially the solid angle defined by the three unit vectors $\mathbf{n}_{i}$ takes the place of the 
Aharanov--Bohm phase in the ordinary Hall effect, so that
$\sigma _{xy}^{a} \sim \text{Im}(t_{12}t_{23}t_{31})\sin (\Omega /2).$

A subtlety that needs commenting on is that upon averaging over the spin orientations, $\Omega $ vanishes since reflection of the vectors 
about any plane containing the direction of magnetization (keeping the magnetization fixed) reverses the sign of  
$\mathbf{n}_{1}\cdot\mathbf{n}_{2}\times\mathbf{n}_{3}.$
However a finite spin-orbit interaction lifts the degeneracy (see e.g. \cite{lyanda2001,millis1999}) by introducing the DM 
interaction of spins in the triad, and this results in a finite anomalous Hall contribution proportional to $\sin(\Omega /2).$ The 
temperature then needs to be smaller than the DM scale to prevent mixing of these states. This mechanism does not require spin splitting of 
charge carrier energies; moreover, it does not exist in a strong exchange mean field as it needs the on-site magnetic moments to be partly disoriented.
This contribution is small near $T_{c}$ where the magnetization is small and steadily increases on cooling, being peaked at magnetization of about 
0.35 of the saturation value~\cite{lyanda2001}. Thus the geometric phase contribution exists only for temperatures below the inflection point (Region IV) 
and at low magnetic fields. Because the above two contributions to AHE are physically different, they can be of different sign.
If we assume that, then the AHE and parametric dependence in Region IV (and others) is easily understood. Since the DM geometric contribution is 
suppressed by a magnetic field, we observe in Fig.~\ref{fig:berry} that the upturn in the AHE seen in Region IV, reverts to the trend in Region III at $B=3$ T. 
The same behavior can also be seen even more clearly in sample M2.

In summary, we have performed detailed measurements and analysis of the anomalous Hall resistance $R_{xy}^{a}$ and
longitudinal resistance $R_{xx}$ of holes in 2D ferromagnetic semiconductor heterostructures that we have earlier established to
have two magnetic transitions corresponding to appearance of local ferromagnetic order and a percolation transition at a lower
temperature $T_{c}.$
At temperatures around the local ferromagnetic
transition, the parametric scaling of $R_{xy}^{a}$ with $R_{xx}$ is observed
to be parabolic, reminiscent of side jump or intrinsic mechanisms in
bulk metallic ferromagnets.
At lower temperatures, $R_{xy}^{a}$ scales linearly
with $R_{xx}$, which has an insulating temperature dependence. At $T_{c},$
the linear scaling persists and the sign of $R_{xy}^{a}$ reverses. We
believe  that the  linear scaling and sign change can both be explained in terms of a spin-dependent hopping transport theory \cite{burkov2003,granov1986}.
At even lower temperatures (around 8--9 K), there is evidence of a new contribution to $R_{xy}^{a}.$ The scaling with $%
R_{xx}$ continues to be linear but this contribution to $R_{xy}^{a}$ has a
positive sign, and it has a strong dependence on the applied magnetic field.
For saturating (and higher) magnetic fields, this contribution becomes
vanishingly small.
We find this behavior to be consistent with a geometric
Berry--Pancharatnam phase picture including the DM interaction. To our understanding, this is the first observation of the geometric phase 
mechanism of the anomalous Hall effect in 2D semiconductor heterostructures. In hindsight, the 2D nature of the structure facilitated the geometric 
phase mechanism because here, unlike the 3D case, there is a window of temperatures above the percolation transition where one has local 
ferromagnetism but the moments are misaligned on longer length scales.

The work was supported by the Russian Ministry of Education and Science (grant 14.613.21.0019), DST (India)(grants RFBR-P-141, RMES-02/14), 
and Russian Science Foundation, projects 14-12-00255 (IVR and NSA) and 14-02-00879 (BAA and LNO).  VT also grateful to DST for a Swarnajayanti 
grant and Argonne Natl. Lab. where a part of the work was carried out. ABD acknowledges a support of RFBR (project 14-02-00586).

\end{document}